\documentclass[10pt,aps,prb,reprint,superscriptaddress,floatfix,longbibliography]{revtex4-2}

\usepackage{physics}
\usepackage{amsmath}
\usepackage{amssymb}
\usepackage{amsthm}
\usepackage[dvipsnames]{xcolor}

\usepackage{amsbsy}
\usepackage{amstext}
\usepackage{graphicx}

\usepackage{color}
\usepackage{mathtools}
\usepackage{subfigure}
\usepackage{multirow}

\usepackage{amsfonts}
\usepackage{dcolumn}
\usepackage{bbold,bm}
\usepackage[bookmarks=false,linkcolor=Orange,urlcolor=MidnightBlue,colorlinks,citecolor=Maroon]{hyperref}

\makeatletter
\newcommand{\A}{\alpha}
\newcommand{\B}{\beta}
\newcommand{\s}{\sigma}
\newcommand{\w}{\omega}
\newcommand{\D}{\Delta}
\newcommand{\kv}{\vb{k}}

\def\E0{\varepsilon_0}

\makeatother
\usepackage{babel}

\begin{abstract}
We theoretically study first and second-order optical responses in a transition metal dichalcogenide monolayer with distinct trivial, nodal, and time-reversal invariant topological superconducting (TRITOPS) phases. We show that the second-order DC response, also known as the photogalvanic response, contains signatures for differentiating these phases while the first-order optical response does not. We find that the high-frequency photogalvanic response is insensitive to the phase of the system, while the low-frequency response exhibits features distinguishing the three phases. At zero doping, corresponding to an electron filling in which the Fermi level lies at nodal points, there are opposite sign zero-frequency divergences in the response when approaching the nodal phase boundaries from the trivial and the TRITOPS phases. In the trivial phase, both the high-frequency and low-frequency response of the system are negative, but in the TRITOPS phase, the low-frequency response becomes positive while the high-frequency response remains negative. Furthermore, since phase transitions are controlled by the Rashba spin-orbit coupling and the ratio of intra-orbital and inter-orbital paring amplitudes, our results not only help distinguish the phases but can also provide an estimate of the pairing amplitudes based on the photogalvanic response of the system.
\end{abstract}

\begin{document}
\allowdisplaybreaks

\title{Nonlinear optical responses in multi-orbital topological superconductors}

\author{Arpit Raj}
\email{raj.a@northeastern.edu}
\affiliation{Department of Physics, Northeastern University, Boston, Massachusetts 02115, USA}

\author{Abigail Postlewaite}
\affiliation{Department of Physics, Northeastern University, Boston, Massachusetts 02115, USA}

\author{Swati Chaudhary}
\affiliation{Department of Physics, The University of Texas at Austin, Austin, Texas 78712, USA}
\affiliation{Department of Physics, Northeastern University, Boston, Massachusetts 02115, USA}
\affiliation{Department of Physics, Massachusetts Institute of Technology, Cambridge, Massachusetts 02139, USA}

\author{Gregory A. Fiete}
\affiliation{Department of Physics, Northeastern University, Boston, Massachusetts 02115, USA}
\affiliation{Department of Physics, Massachusetts Institute of Technology, Cambridge, Massachusetts 02139, USA}

% ----------------------------------------------------------------------------------------
\maketitle

\section{Introduction}

The second-order nonlinear optical response serves as a highly effective tool for probing symmetry-broken states~\cite{belinicher1980photogalvanic,orenstein2021topology}. The second-order DC response to an alternating electric field, also known as the photogalvanic effect, has taken central stage recently~\cite{morimoto2023geometric,ma2021topology,ma2023photocurrent,Juan2017,Bhalla2020,cook2017design,Watanabe2021,Parker19,wu2017giant,Chan2017,Konig2017,osterhoudt2019colossal,yuan2014generation,ma2017direct,gao2020chiral}. Photogalvanic effects have been widely employed for probing the symmetries of quantum phases, as well as non-trivial quantum geometries of electronic bands~\cite{morimoto2016topological, Ahn2020PRX,ahn2022riemannian,sipe2000,Holder2020,nagaosa2017concept,Baltz1981,Kraut1979,Chaudhary2022,Kaplan2022TBG,Gao2020,raj2023photogalvanic}. Recent theory works have established that nonlinear responses are interesting for noncentrosymmetric superconductors~\cite{xu2019,Nakamura2020,Hikaru2022, tanaka2023} and can aid in the characterization of the phase (topological or trivial) and order parameter symmetry of experimentally identified superconductors.

While the linear optical conductivity has long been used to probe quantities like spectral weight transfer, the nature of the superconducting state, and magnitude of the superconducting gap~\cite{Tinkham1956,Tinkham1968, ahn2021theory, Papaj2022}, recently experiments have also studied nonlinear optical properties of cuprate superconductors, revealing the nature of the broken symmetries in the pseudogap phase~\cite{zhao2017global}. It was proposed in Ref.~\cite{xu2019} that the signatures of inversion-breaking superconductivity are much stronger in second-order optical effects than in the linear optical conductivity, and can persist over a relatively wide range of temperatures making the nonlinear response an important quantity to study.

The absence of inversion symmetry in noncentrosymmetric superconductors allows for the coexistence of opposite parity pairing channels leading to a mixed-parity order parameter~\cite{Gorkov2001,bauer2012non, yip2014noncentrosymmetric,smidman2017superconductivity,fischer2023superconductivity,Nogaki2022,sonowal2023second}. A mixed parity order parameter can lead to the emergence of exotic superconducting effects, such as the nonreciprocal Meissner effect~\cite{Hikaru2022Meissner}, finite momentum pairing states~\cite{saito2016superconductivity,wan2023orbital}, topological superconductivity~\cite{zhang2022topological,gao2022topological} and helical superconductivity~\cite{yanase2008helical}. Monolayer and few layer transitions metal dichalcogenides (TMDs) are known for their  wealth of electronic and magnetic phases~\cite{wang2012electronics,manzeli20172d,yang2017structural}. At low temperatures, TMDs constitute one class of superconductors that have been identified as highly suitable candidates for a mixed parity pairing potential and exotic superconductivity~\cite{Castro2001, lu2015evidence, Liu2017PRL, zhang2020superconductivity, Lane2022}.

Layered TMDs are proving to be a highly versatile platform for studying unconventional superconductivity~\cite{Yuan2014, Mockli2018, yuan2019evidence, nayak2021evidence, Guinea2013Mos2, hsu2017topological}. TMD monolayers MX$_2$ (M=Mo, W, Nb, Ta, X= S, Se, Te), lack an inversion center and have significant electronic correlations and spin-orbit coupling (SOC) which makes them ideal candidates for topological superconductivity and unconventional pairing~\cite{de2018tuning, Yuan2014, hsu2017topological}. The unconventional superconductivity in group VI TMDs is usually induced by external factors like ionic gating, doping, and intercalation ~\cite{yang2017structural,  saito2016superconductivity, lu2015evidence, li2021observation, hsu2017topological, Oiwa2018} with the exception of 2M-WS$_2$, which is an intrinsic topological superconductor~\cite{li2021observation}.  Group VI layered TMDs NbSe$_2$, TaS$_2$ are known for naturally occurring Ising superconductivity~\cite{xi2016ising, xing2017ising, yang2018prb, lian2018unveiling, Lian2023, Lian2021, Darshana2020, Chen2019IsingTaS2, hamill2021two} which shows remarkable stability to in-plane magnetic fields.

Group VI layered TMDs have gained significant attention as they can exhibit many exotic superconducting features such as nodal superconductivity~\cite{vavno2021evidence}, collective Leggett modes~\cite{wan2022observation} and topological boundary modes~\cite{nayak2021evidence}. In addition to Ising SOC, the multi-band and mixed-parity nature of the pairing terms~\cite{horhold2023two, Cho2022nodal, margalit2021} also endows the superconductivity with many intriguing features in these materials. All these factors lead to a rich phase diagram where superconducting phases with different topological features can be obtained by tuning the in-plane magnetic field, Rashba SOC, and inter-orbital pairing term~\cite{ he2018magnetic,Daniel2020,margalit2021,bao2023magnetic,silber2022chiral}. Thus, this class of TMDs is important for the investigation of topological superconductivity.  It is crucial to characterize the different phases which can be obtained in this class of superconductors and predict what their signatures will be in different types of measurements. Given the controversy around claims of topological superconductivity, it is important to correlate measurements of different types, each of which can provide evidence either for or against topological superconductivity.

Atomically thin TaS$_2$ has recently emerged as a compelling candidate with significant potential for the realization of topological superconductivity~\cite{Galvis2014, Lian2023, navarro2016enhanced}. Ising superconductivity in this material can be enhanced through electron doping or by reducing the number of atomic layers~\cite{navarro2016enhanced}, a phenomenon often ascribed to the suppression of a charge-density wave~\cite{Lian2023}. Possible signatures of  topological superconductivity have been observed in 2H-TaS$_2$ which displays a zero-bias conductance peak in detached flakes of superconducting samples~\cite{Galvis2014}, and most recently in 4Hb-TaS$_2$ (which consists of alternately stacked 1H-TaS$_2$ and 1T-TaS$_2$) which hosts one-dimensional boundary modes~\cite{nayak2021evidence}.
The inter-orbital pairing channel is crucial for non-trivial topology in layered TaS$_2$ and by changing the strength of the inter-orbital pairing term in 1H-TaS$_2$, the system can be driven from a conventional superconductor to a nodal superconductor, and then to a fully gapped time-reversal invariant topological superconductor (TRITOPS)~\cite{margalit2021}.

In this work, we explore the possibility of using the nonlinear optical response (NLOR) to distinguish different topological phases in 1H-TaS$_2$. We study the first (for a linear response comparison) and second-order DC conductivity in three different phases: (i) trivial, (ii) nodal, and (iii) TRITOPS.  In 1H-TaS$_2$, the topological features of the superconducting phase are determined by the Rashba SOC and the ratio of on-site intra-orbital and inter-orbital superconducting pairing terms~\cite{margalit2021} which have opposite parity. It has been shown that photocurrents can carry strong signatures of mixed-parity and multi-band pairing terms~\cite{Hikaru2022,tanaka2023, ahn2021theory}. Multi-band systems also allow for some intrinsic optical excitations which are absent in single-band models~\cite{ahn2021theory} and hence may manifest strongly in non-linear optical responses as well. The multi-orbital nature of superconductivity pairing also leads to non-trivial quantum geometry engendering features like flat-band superconductivity~\cite{Torma2023,chen2023towards,hu2023anomalous,Torma19,Herzog2022,Xie2020,verma2021optical}. Similar quantum geometric aspects also lead to unique signatures in light-matter coupling based processes~\cite{Topp2021} and in particular manifest very strongly in non-linear optical responses~\cite{orenstein2021topology,ma2021topology,ahn2022riemannian,nagaosa2017concept,morimoto2023geometric}.

\begin{figure}
    \centering
    \includegraphics[scale=0.37]{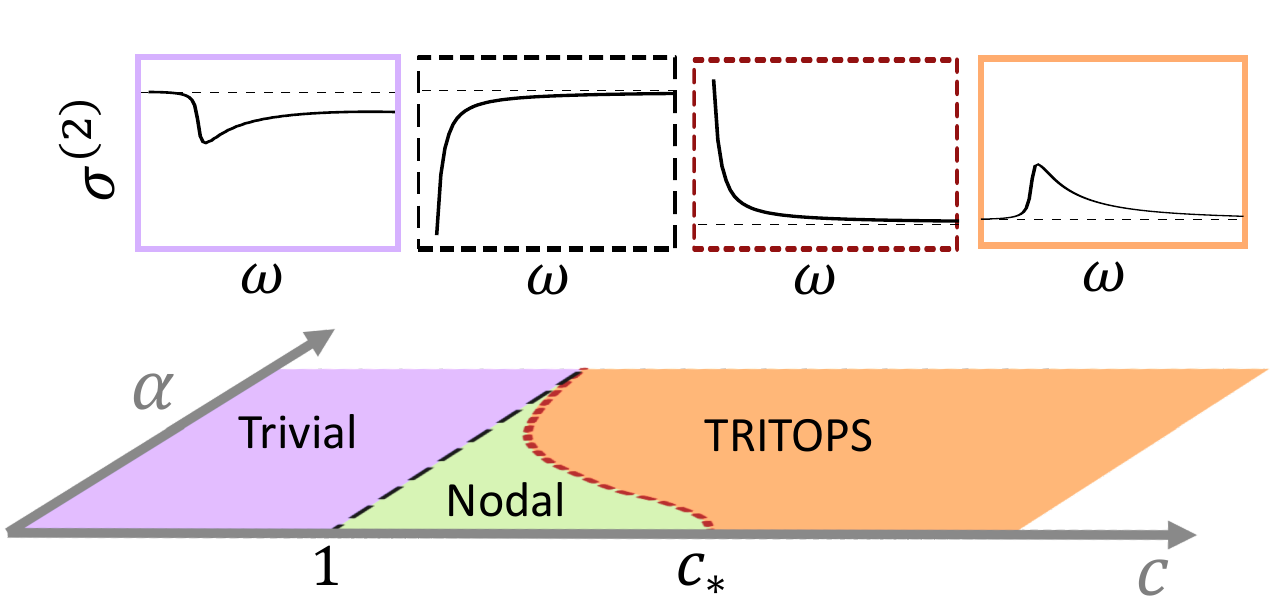}
    \caption{Schematic representation of second-order conductivity in different superconducting phases of  4Hb-TaS$_2$ for low frequencies (close to the superconducting pairing strength). The two gapped phases--shown in purple (trivial) and orange (TRITOPS)--have opposite sign of the second-order conductivity. The magnitude of the response starts to diverge when the nodal phase is approached from either side, as shown by the results boxed with dashed lines. On the other hand, the high frequency behavior (not depicted here) remains the same in all phases.  See Fig.~\ref{fig:phases} for a description of the phase diagram in the lower portion of the figure.}
    \label{fig:schematic}
\end{figure}

Motivated by these works, we calculate the second order DC response of 1H-TaS$_2$ for linearly polarized light. We find that the low-frequency behavior exhibits distinguishing features for three superconducting phases and can serve as a reliable probe to characterize the nature of superconducting state in this exciting Ising superconductor. A schematic for the distinct types second-order response and the associated superconducting phases is shown in Fig.~\ref{fig:schematic}. 

Our paper is organized as follows. In Sec.~\ref{sec:Model} we introduce our theoretical model. We describe the phases of the model, methods to characterize their topological character, and present a global phase diagram.  We also give the expressions used to compute the nonlinear (second-order) optical response. In Sec.~\ref{sec:Results} we describe the main results of our work showing that it is the low-frequency response that characterizes the superconducting state. Finally, in Sec.~\ref{sec:Conclusions} we present the central conclusions of our work and an outlook for issues for future study. Various technical details of the calculations appear in the appendices.

% ----------------------------------------------------------------------------------------
\section{Model}
\label{sec:Model}
\subsection{Hamiltonian}
We consider a two-dimensional (2D) TMD monolayer of tantalum disulfide (1H-TaS$_2$) on a substrate, which breaks inversion and mirror symmetry. We use the density functional theory based six-band tight-binding model provided in Ref.~\cite{margalit2021}. The system can be described by three orbital degrees of freedom $d_{z^2}$, $d_{x^2-y^2}$, and $d_{xy}$, along with two spin degrees of freedom. The substrate plays an important role as it breaks the mirror symmetry of the $z$-axis ($M_z$)  but  mirror symmetry about the $x$-axis ($M_x$) and symmetry under rotation by 120$^\circ$ about the $z$-axis ($C_3$) remain.

The single-particle Hamiltonian $H_0(\kv)$, Eq.(A1), describing this system is formulated using a tight-binding model considering on-site pairings and up to third-nearest-neighbor hopping terms. The details of the Hamiltonian and the hopping parameters~\cite{margalit2021} are given in Appendix~\ref{appendix:model}. However, we also consider the case of chemical potential $\mu=0$ in addition to $\mu=-50$ meV going beyond the regime studied in Ref.~\cite{margalit2021}.  (We note that Ref.~\cite{margalit2021} did not study any aspects of the optical response of the model.)   

One can directly add $M_z$ symmetry breaking terms to $H_0(\kv)$, such as a Rashba spin-orbit coupling term, 
\begin{align}
    H_{\rm{Rashba}}(\kv) &= i\A \sum_{j=1}^6 (R_{j}^x\s_y-R_{j}^y\s_x)e^ {i\mathbf{R}_j\cdot{\kv}} \otimes \mathbb{I}_3, \label{eq:rashba}
\end{align}
where $\s_x$, $\s_y$ are Pauli matrices, $\mathbf{R}_j$ are the lattice vectors for nearest-neighbor sites, $\mathbb{I}_3$ is the $3\times 3$ identity matrix, and $\A$ is a constant describing the strength of the Rashba term. This time-reversal invariant term plays an important role in controlling the phase of the superconducting Hamiltonian.

Superconductivity is incorporated via the Boguliubov-de Gennes formalism, with the 12 x 12 superconducting Hamiltonian given by, 
\begin{align} \label{eq:HBdG}
    H_{SC}(\mathbf{k}) &= \mqty[H_0(\mathbf{k}) & \D \\
                                \D ^\dagger & -H_0(-\mathbf{k})^T],
\end{align}
which is written in the Nambu basis $\Psi^{\dagger}_{\kv} = (\psi^{\dagger}_{\kv},\psi^T_{-\kv})$ with
\begin{align}
    \psi^T_\mathbf{k} &= (d_{z^2,\uparrow}, d_{xy,\uparrow},d_{x^2-y^2,\uparrow},d_{z^2,\downarrow}, d_{xy,\downarrow},d_{x^2-y^2,\downarrow}),
\end{align}
where $d_{\nu,\s}(\kv)$ are annihilation operators acting on electrons with spin $\s$ in orbital $\nu$. In Eq.~\eqref{eq:HBdG}, $\D$ is a momentum-{\it independent} pairing matrix determined from symmetries of the model.

The anticommutativity of fermions generally requires $\D(\kv) = -\D^T(-\kv)$. In a momentum-independent matrix, we thus have $\D = -\D^T$. In a single-orbital system this requirement eliminates all but the trivial $i\s_y$ term from the pairing matrix. However, the additional degrees of freedom contained within a multi-orbital system allow other terms, provided the pairing matrix remains anti-symmetric in the orbital degree of freedom. The most general form of the pairing matrix consistent with the symmetries of the system is given by~\cite{margalit2021}:
\begin{align} \label{eq:pairing-matrix}
\D &=\mqty[0 & \D_4 & i\D_4 & \D_1 & 0 & 0 \\
         -\D_4 & 0 & 0 & 0 & \D_2 & i\D_3 \\
         -i\D_4 & 0 & 0 & 0  & -i\D_3 & \D_2 \\
         -\D_1 & 0 & 0 & 0 & \D_4 & -i\D_4 \\
         0 & -\D_2 & i\D_3 & -\D_4 & 0 & 0 \\
         0 & -i\D_3 & -\D_2 & i\D_4 & 0 & 0],
\end{align}
where $\D_{1,2,3,4}$ are real parameters. Here, $\D_1$ describes intra-orbital singlet pairing in the $d_{z^2}$ orbital while $\D_2$ describes intra-orbital singlet pairing within the $d_{x^2-y^2}$ and $d_{xy}$ (in-plane) orbitals. Similarly, $\D_3$ gives the inter-orbital triplet pairing of the $d_{xy}$ and $d_{x^2-y^2}$ orbitals.  The parameter $\D_4$ is also an inter-orbital triplet term and gives the pairing of same-spin states. In our numerical analysis of the nonlinear optical response, we set $\D_2 = \D_1$ and $\D_3 = 0$. The ratio $c = \frac{\D_4}{\zeta\D_1}$, where $\zeta$ is a model-dependent parameter, is an important quantity controlling the phase of the model. The pair of quantities $(c,\A)$ can be varied to drive the system across the different phases as discussed in Fig.~\ref{fig:phases} and Sec.~\ref{subsec:Phases}.  This model has successfully explained the crystal orientation dependent local density of states of edge modes observed in scanning tunneling microscope experiments \cite{nayak2021evidence}.

% ----------------------------------------------------------------------------------------
\subsection{Phases and topology}
\label{subsec:Phases}
The model described with the pairing matrix given in Eq.~\eqref{eq:pairing-matrix} shows three distinct phases - trivial, nodal, and TRITOPS. The trivial and TRITOPS phases are gapped, and based on the symmetry class of the system can be distinguished with a $\mathbb{Z}_2$ invariant~\cite{margalit2021}.  One can compute the invariant by putting $H_{SC}$ in an off-diagonal block form. To do this we first change the basis of the hole block of the BdG spinor, following the work of Ref.~\cite{margalit2021}, 
\begin{align}
 \widetilde{\Psi} &= \mqty[\mathbb{I}_2\otimes\mathbb{I}_3 && 0 \\
                            0 && i\s_y\otimes\mathbb{I}_3]
                    \mqty[\psi_\mathbf{k} \\
                          \psi_\mathbf{-k}^\dagger].
\end{align}
The Hamiltonian in the transformed basis, $\widetilde{H}_{SC}(\kv)$, can then be expressed in the off-diagonal form,
\begin{equation}
    e^{i\frac{\pi}{4}\tau_x}\widetilde{H}_{SC}(\mathbf{k})e^{-i\frac{\pi}{4}\tau_x} =
    \begin{bmatrix}
        0 && Q_\mathbf{k} \\
        Q^{\dagger}_\mathbf{k} && 0
    \end{bmatrix},
\end{equation}
where $\tau_x$ is a Pauli matrix in the particle-hole subspace. The matrix $Q_{\kv}$ is related to the single particle Hamiltonian and the paring matrix as,
\begin{align}
    iQ_{\kv} &= H_0(\kv)+iU_\mathcal{T}\D^\dagger,
\end{align}
where $U_\mathcal{T} = i\s_y\otimes\mathbb{I}_3$. With $Q_{\kv}$, the $\mathbb{Z}_2$ invariant can be explicitly calculated following Ref.~\cite{qi2010topological}. However, in the weak pairing limit, when $\D$ is tiny compared to the energy separation between bands, one can calculate the topological invariant by simply looking at the sign of the effective pairing $\delta_{n,\kv}$ for bands crossing the Fermi level. The effective pairing is defined as,
\begin{equation}
    \delta_{n,\kv} = \bra{n,\kv}U_\mathcal{T}\D^\dagger\ket{n,\kv},
\end{equation}
where $\ket{n,\kv}$ is an eigenstate of $H_0(\kv)$. In this limit, the system is topological if there are an odd number of Fermi pockets, each enclosing one TRIM point and with a negative $\delta_{n,\kv}$~\cite{qi2010topological}. Since this invariant is only well defined when the system is gapped, in the nodal phase it is sometimes useful to look at the characteristic angle, $\theta_{\kv}$, defined as, 
\begin{equation}
    \theta_{\kv} = \arg(\det Q_{\kv}),
\end{equation}
whose winding around a node gives its topological charge.

\begin{figure*}[ht!]
    \includegraphics[scale=0.58]{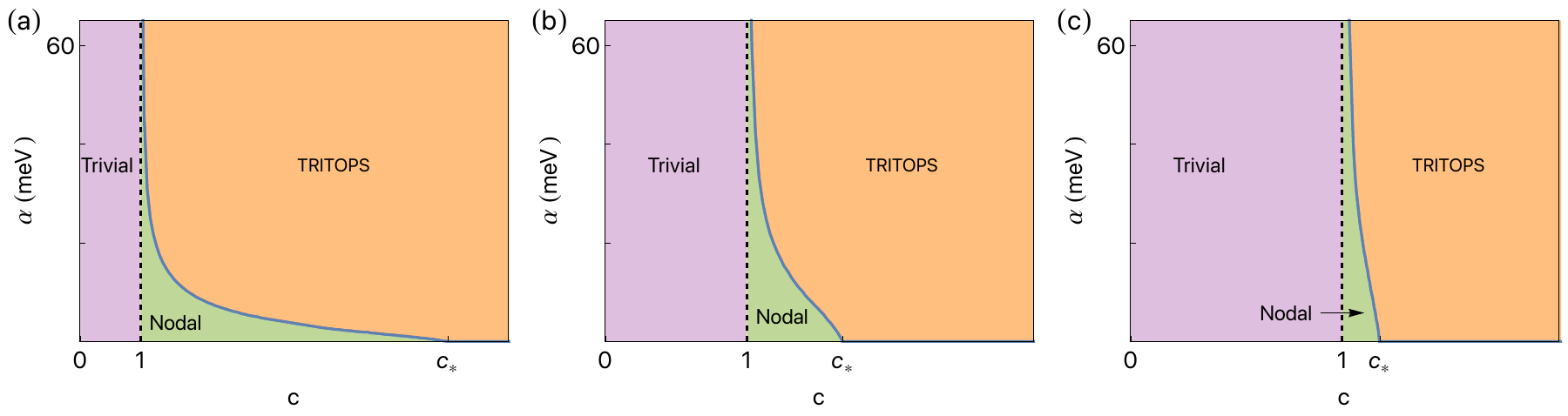}
    \caption{Phase diagram in the $\A$-$c$ plane for $\mu=0$ and (a) $\D_1=10$ meV, $c_*=6.054$, (b) $\D_1=45$ meV, $c_*=1.665$, (c) $\D_1=100$ meV, $c_*=1.1705$.  Smaller $\D_1$ values (which set the scale of the superconducting transition temperature) are closer to experimental values for superconducting TMDS. For smaller $\D_1$ the $c$ values over which the nodal phase is possible is enlarged, but the phase is more easily destroyed by spin-orbit coupling, whose strength is given by $\A$.}
    \label{fig:phases}
\end{figure*}
\begin{figure*}[ht!]
    \includegraphics[scale=0.53]{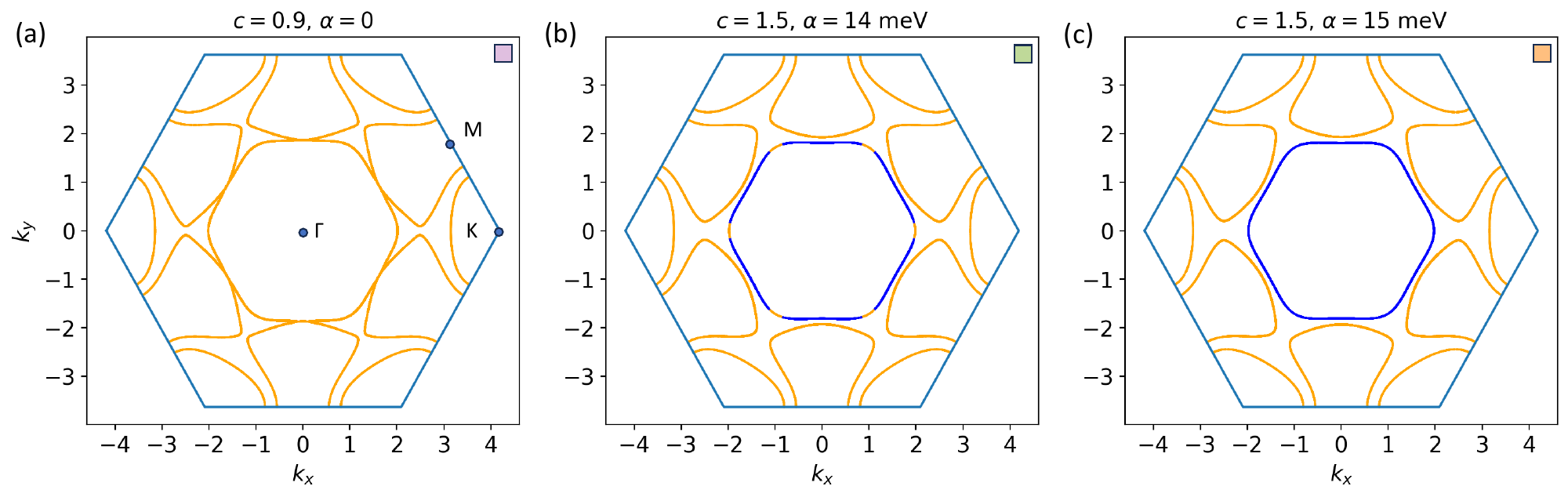}
    \caption{Sign of the effective paring for the (a) trivial, (b) nodal, (c) TRITOPS phases, where blue (orange) corresponds to $-1\,(+1)$. We used $\D_1=10\,\rm{meV}$ and $\mu=-50\,\rm{meV}$ for these plots. Note that for $c=1.5$, $\A_c=14.58$ meV.  The small colored square box in the upper right corner of each sub-panel in the figure corresponds to the phase of the same color in Fig.~\ref{fig:phases}. }
    \label{fig:effpairing}
\end{figure*}
As seen in Fig.~\ref{fig:phases}, the trivial to nodal phase transition can only be driven with $c=\frac{\Delta_4}{\zeta \Delta_1}$. The parameter $\zeta$ used in the definition is chosen such that $c = 1$ is the transition point from the trivial to the nodal phase at $\mu=0$. When entering the nodal phase, the nodes appear on the innermost Fermi surface (of the normal Hamiltonian) along the $\Gamma-\rm{M}$ lines. (See Fig.~\ref{fig:effpairing}.) When $c$ is increased further, each node splits into two and start to move away from their origin point (but stay on the same Fermi surface due to time-reversal symmetry~\cite{sato2006, berg2008, beri2010}). At $c=c_*$, oppositely charged nodes merge on that Fermi surface, along the $\Gamma-\rm{K}$ lines, marking a transition to the gapped TRITOPS phase. We should note that once in the nodal phase ($1<c<c_*$), it is possible to drive the  system into the TRITOPS phase by turning on $\A$. Increasing $\A$ causes the nodes to move  along the Fermi surface as before and finally merge along the $\Gamma-\rm{K}$ lines at $\A=\A_c$. In fact, $\D_1$ and $c_*$ are inversely proportional, making the $\A$ driven nodal to TRITOPS phase transition more accessible for values of pairing amplitudes used for 1H-TMDs in Ref.~\cite{margalit2021}.

With this picture of node creation and annihilation in mind, one can determine the two phase boundaries by looking at the gap closing along the $\Gamma$-$\rm{M}$ and the $\Gamma$-$\rm{K}$ lines. Phase diagrams obtained using this approach are shown in Fig.~\ref{fig:phases} for three different values of the intra-orbital pairing term $\D_1$. These phase transitions can also be confirmed by looking at the sign of the effective paring on the Fermi surface across the phase boundaries, as shown in Fig.~\ref{fig:effpairing}. In the trivial phase shown in Fig.~\ref{fig:effpairing}~(a), the Fermi surface around all four TRIM points has a positive sign for the effective pairing. On the other hand, for the TRITOPS phase shown in Fig.~\ref{fig:effpairing}~(c), the Fermi pocket around the $\Gamma$ point acquires a negative sign which indicates the non-trivial nature of this phase. Some of the phase transition points considered in our nonlinear conductivity calculations are given in Tables~\ref{tab:zeromu} and \ref{tab:finitmu} for the chemical potentials $\mu=0$ and $\mu=-50$ meV, respectively. Note that for $\mu=-50$ meV the trivial to nodal phase transition point does not change significantly from $c=1$.
\begin{table}[ht!]
 \caption{Various nodal to TRITOPS phase transition points for $\mu=0$ for different values of $\D_1$. Note that at zero doping, the trivial to nodal transition is always at $c=1$ by construction.}
\begin{ruledtabular}
\begin{tabular}{c|c|c|c}
    $\D_1$ (meV) & $\zeta$ & $c$  & $\A_c\,\rm{(meV)}$ \\
    \hline
    \multirow{2}{*}{10} & \multirow{2}{*}{1.171478265857} & 1.5 & 13.31677465 \\\cline{3-4}
    &  & 6.05433710 & 0 \\
    \hline
    \multirow{2}{*}{45} & \multirow{2}{*}{1.170919255973} & 1.3 & 13.452 \\\cline{3-4}
    &  & 1.66527725 & 0\\
    \hline
    100 & 1.168647137 & 1.17054149 & 0 
\end{tabular}
\end{ruledtabular}
\label{tab:zeromu}
\end{table}
\begin{table}[ht!]
\caption{Various phase transition points for $\mu=-50$ meV for different $\D_1$ (respective $\zeta$ value kept same as for $\mu=0$).}
\begin{ruledtabular}
\begin{tabular}{c|c|c|c}
    $\D_1$ (meV) & Transition & $c$ & $\A_c\,\rm{(meV)}$ \\
    \hline
    \multirow{2}{*}{10} & Trivial $\rightarrow$ Nodal & 1.0087 & - \\\cline{2-4}
    & Nodal $\rightarrow$ TRITOPS & 1.5 & 14.58 \\
    \hline
    \multirow{3}{*}{45} & Trivial $\rightarrow$ Nodal & 1.00194861 & - \\\cline{2-4}
    & \multirow{2}{*}{Nodal $\rightarrow$ TRITOPS} & 1.3 & 15.62597 \\
    & & 1.7917638 & 0 
\end{tabular}
\end{ruledtabular}
\label{tab:finitmu}
\end{table}

% ----------------------------------------------------------------------------------------
\subsection{Nonlinear Optical Response}
We study the second-order DC response, also known as the photogalvanic effect, following Ref.~\cite{Hikaru2022, tanaka2023}. Using the expression for the second-order conductivity,
\begin{widetext}
\begin{align}
\begin{split}
    \s^{\A\B\gamma}(\widetilde{\w};\w_1,\w_2) &= \int_{\rm{FBZ}} \frac{\dd[2]{k}}{(2\pi)^2}\, \frac{1}{2(i\w_1-\eta)(i\w_2-\eta)} \Bigg[\sum_a \frac{1}{2}J_{aa}^{\A\B\gamma}f_a + \sum_{a,b} \frac{1}{2}\Bigg( \frac{J_{ab}^{\A\B}J_{ba}^\gamma f_{ab}}{\w_2+i\eta -E_{ba}} + \frac{J_{ab}^{\A\gamma}J_{ba}^\B f_{ab}}{\w_1+i\eta -E_{ba}}\Bigg)\\
    &\quad + \sum_{a,b}\frac{1}{2} \frac{J_{ab}^\A J^{\B\gamma}_{ba}f_{ab}}{\widetilde{\w} + 2i\eta -E_{ba}} + \sum_{a,b,c} \frac{1}{2}\frac{J_{ab}^\A}{\widetilde{\w} + 2i\eta-E_{ba}}\Bigg(\frac{J_{bc}^\B J_{ca}^\gamma f_{ac}}{\w_2+i\eta-E_{ca}}-\frac{J_{ca}^\B J_{bc}^\gamma f_{cb}}{\w_2 + i\eta -E_{bc}}\Bigg)\\
    &\quad + \sum_{a,b,c} \frac{1}{2}\frac{J_{ab}^\A}{\widetilde{\w} + 2i\eta-E_{ba}}\Bigg(\frac{J_{bc}^\gamma J_{ca}^\B f_{ac}}{\w_1+i\eta-E_{ca}}-\frac{J_{ca}^\gamma J_{bc}^\B f_{cb}}{\w_1 + i\eta -E_{bc}}\Bigg)\Bigg],
\end{split}\label{eq:sigma2}
\end{align}
\end{widetext}
where $\widetilde{\w}=\w_1+\w_2$, we calculate the photogalvanic response by computing the DC conductivities $\s^{\A\B\gamma}(0;\w,-\w)$. In Fig.~\ref{fig:sigmayyy1}-Fig.~\ref{fig:sigmayyy3} we plot $\s^{\A\B\gamma}(0;\w,-\w)$ as a function of $\omega$.  Here, $J_{ab}^{\A}$, $J_{ab}^{\A\B}$, and $J_{ab}^{\A\B\gamma}$ are matrix elements of the generalized current operator $J$ defined as~\cite{Hikaru2022, tanaka2023},
\begin{align}
    J^{\A_1,\A_2,\ldots,\A_n} &= (-1)^n \eval{ \frac{\partial^n H_{SC}(\kv,\boldsymbol{\lambda})}{\partial\lambda^{\A_1} \partial\lambda^{\A_2} \ldots \partial\lambda^{\A_n}}}_{\boldsymbol{\lambda}=0},
\end{align}
where,
\begin{align}
    H_{SC}(\kv,\boldsymbol{\lambda}) &= 
    \mqty [ H_0(\kv-\boldsymbol{\lambda}) & \D \\
            \D^\dagger & -H_0(-\kv-\boldsymbol{\lambda})^T].
\end{align}

To numerically evaluate Eq.~\eqref{eq:sigma2}, the integral is converted to a sum over discrete $\kv$-points in the first Brillouin zone of the system. Latin indices label the eigenvectors of $H_{SC}(\kv)$ with $E_{ab} = E_a - E_b$ and $f_{ab} = f_a - f_b$, where $f_a$ refers to the Fermi-Dirac distribution function $f_a = 1/(1+e^{E_a/k_BT})$. We take $T=10^{-4}$ K and a small phenomenological scattering rate $\eta = 5\times10^{-4}$ eV for $\D_1=100,\,45$ meV and $\eta = 1.5\times10^{-4}$ eV for $\D_1=10$ meV. Note that these pairing amplitudes are about $10-100$ times larger that the one given in Ref.~\cite{margalit2021}, which are around $1\,\rm{meV}$. Working with pairings on the order of $1\,\rm{meV}$ results in unreliable numerical results unless one uses an extremely fine $k$-grid, which in turn leads to high computational time. To overcome this, we work with large pairing amplitudes instead and then show that decreasing them has a clear trend in terms of certain features that are of interest to us, such as the low-frequency behavior.  One can then safely extrapolate the trends down to pairing on the order of meV. (Such trends are also reflected in the critical boundaries of the phase diagrams themselves, as seen in Fig.~\ref{fig:phases} and Tables~\ref{tab:zeromu}, \ref{tab:finitmu}.)  The sign of the divergence in the low-frequency regime is what provides the strongest fingerprint of  different superconducting phases.

For comparison we also examine the first-order conductivity of the system, given by~\cite{Hikaru2022},
\begin{align}
    \s^{\A\B}(\w) = \frac{i}{2(\w + i\eta)}\sum_{a,b} \Bigg(\frac{J_{ab}^\A J_{ba}^\B f_{ab}}{\w+i\eta -E_{ba}} + J_{ab}^{\A\B} f_a \delta_{ab} \Bigg ), \label{eq:sigma1}
\end{align}
for signatures of the topological phases of the system.  We find that the second order response in Eq.~\eqref{eq:sigma2} reflects the superconducting phase and transitions between phases while the first order response in Eq.~\eqref{eq:sigma1} does not. This physical result provides an excellent example of additional physics being obtained though higher order responses.

\begin{figure*}[ht!]
    \includegraphics[scale=0.58]{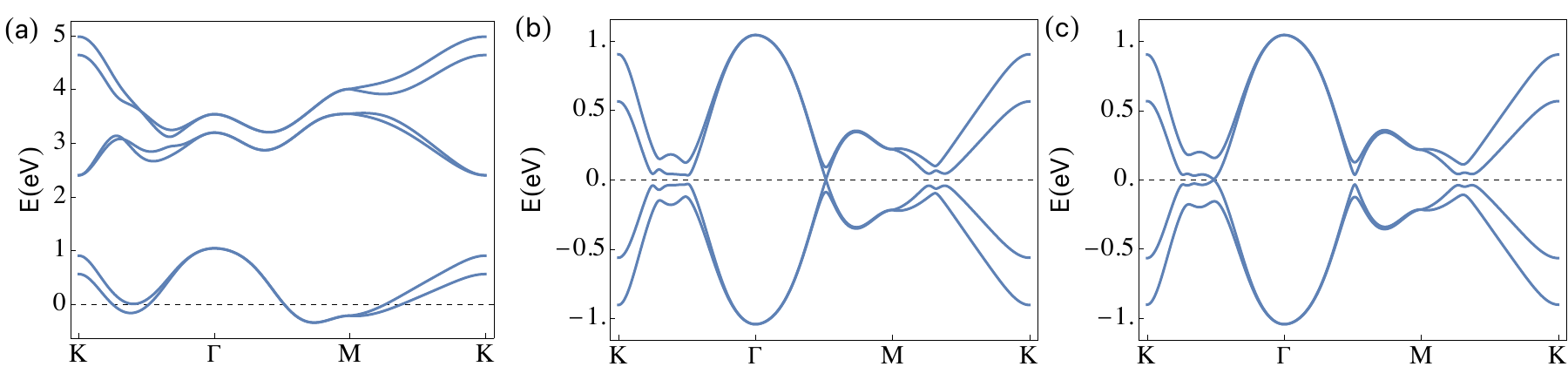}
    \caption{Band structure for the single-particle Hamiltonian $H_0$ without Rashba SOC at $\mu=-50$ meV is shown in (a). The energy bands close to the Fermi energy for the Boguliubov-de Gennes Hamiltonian $H_{SC}$ with $\D_1=45$ meV for (b) $c=1.0019486$, $\A=0$ and (c) $c=1.7917638$, $\A=0$ are also shown for the same chemical potential. One can see a gap closing along the $\Gamma-$M line at the trivial-nodal and along the $\Gamma-$K line at the nodal-TRITOPS phase transition points in (b) and (c), respectively.}
    \label{fig:bands}
\end{figure*}
\begin{figure*}[ht!]
    \centering   \includegraphics[trim=0.5cm 0.5cm 0cm 0.5cm, clip,scale=0.58]{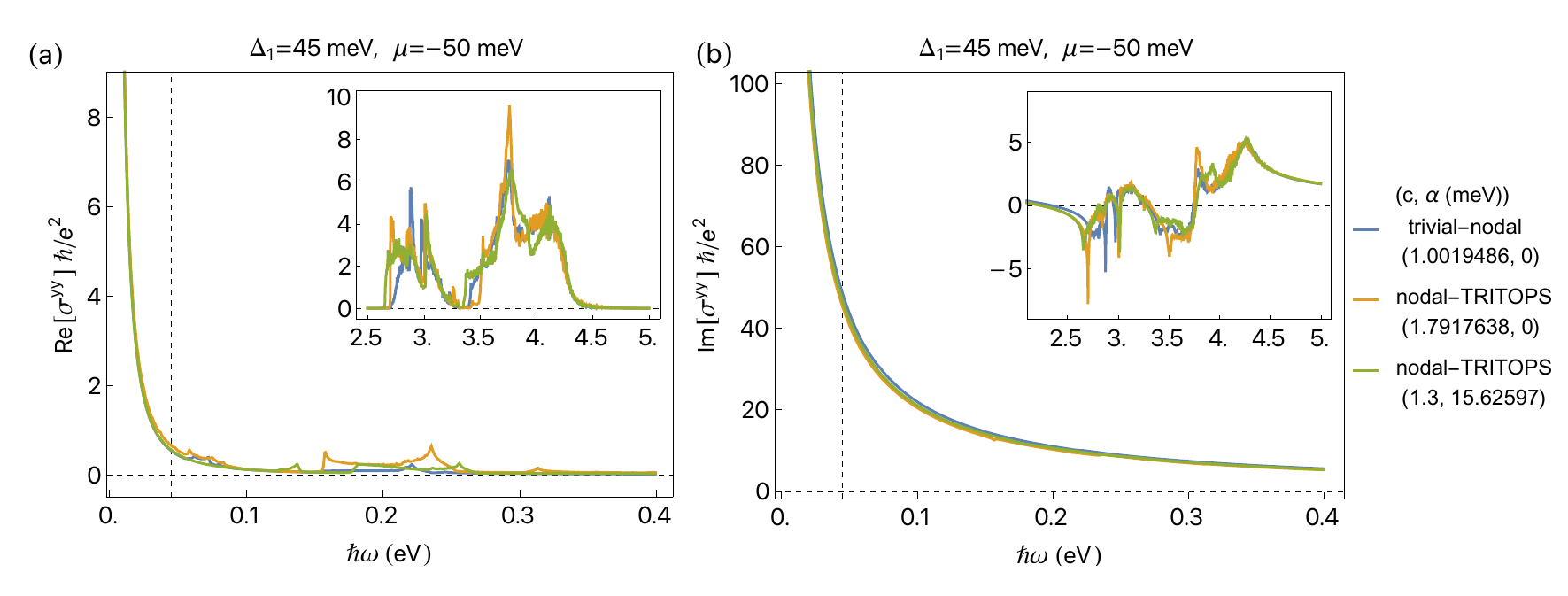}
    \caption{The real and imaginary part of first order conductivity $\s^{xx}(\w)=\s^{yy}(\w)$ for $\D_1=45\,\rm{meV}$, and $\mu=-50$ meV. Both the low and high frequency response remain unchanged across the trivial-nodal and nodal-TRITOPS phase transitions points.}
    \label{fig:sigmayy}
\end{figure*}
% ----------------------------------------------------------------------------------------
\section{Results}
\label{sec:Results}

\begin{figure*}[ht!]
    \includegraphics[scale=0.58]{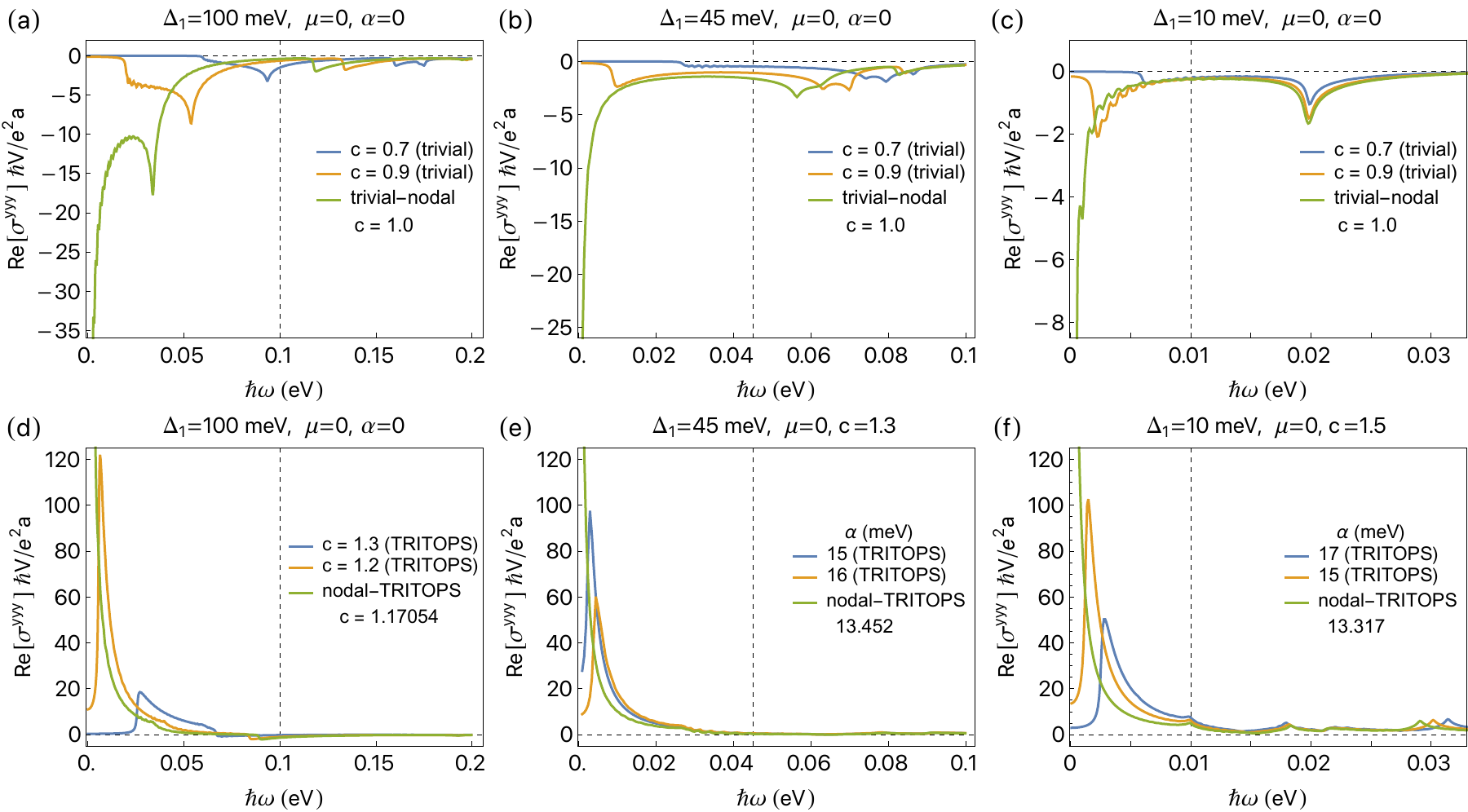}
    \caption{$\rm{Re}[\s^{yyy}(\w)]$ for different values of $\D_1$, $c$, and $\A$ for $\mu=0$.
    (a-c) shows the behavior when approaching the trivial-nodal phase boundary whereas (d-f) captures the behavior past the nodal-TRITOPS phase boundary. Close to phase boundary, sign of the low-frequency photogalvanic response in the trivial and TRITOPS phase matches with that of the divergence. Note that we chose to drive the nodal to TRITOPS phase transition with $c$ in (d) and $\A$ in (e,f), showing no qualitative difference. The vertical dashed line in each figure indicates the value of $\Delta_1$. The factor $a$ appearing in the $y$-axis label is the lattice constant for 1H-Ta$\rm{S}_2$ monolayer.}
    \label{fig:sigmayyy1}
\end{figure*}
\begin{figure*}[ht!]   \includegraphics[scale=0.58]{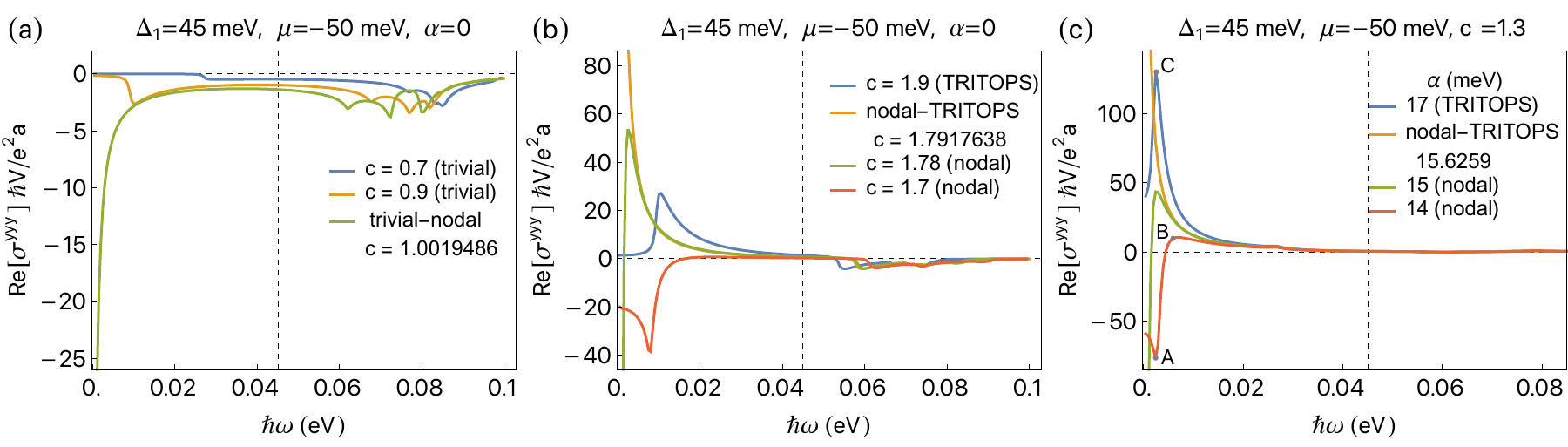}
    \caption{$\rm{Re}[\s^{yyy}(\w)]$ for $\D_1=45\,\rm{meV}$, and $\mu=-50\,\rm{meV}$. (a) shows the behavior when approaching the trivial-nodal phase boundary. (b) and (c) capture the behavior across the nodal-TRITOPS phase boundary for $c$ and $\A$ driven transitions, respectively. Again, we see no qualitative difference between the two routes. The vertical dashed line indicates the value of $\Delta_1$.}
    \label{fig:sigmayyy2}
\end{figure*}
\begin{figure*}[ht!]
\includegraphics[scale=0.58]{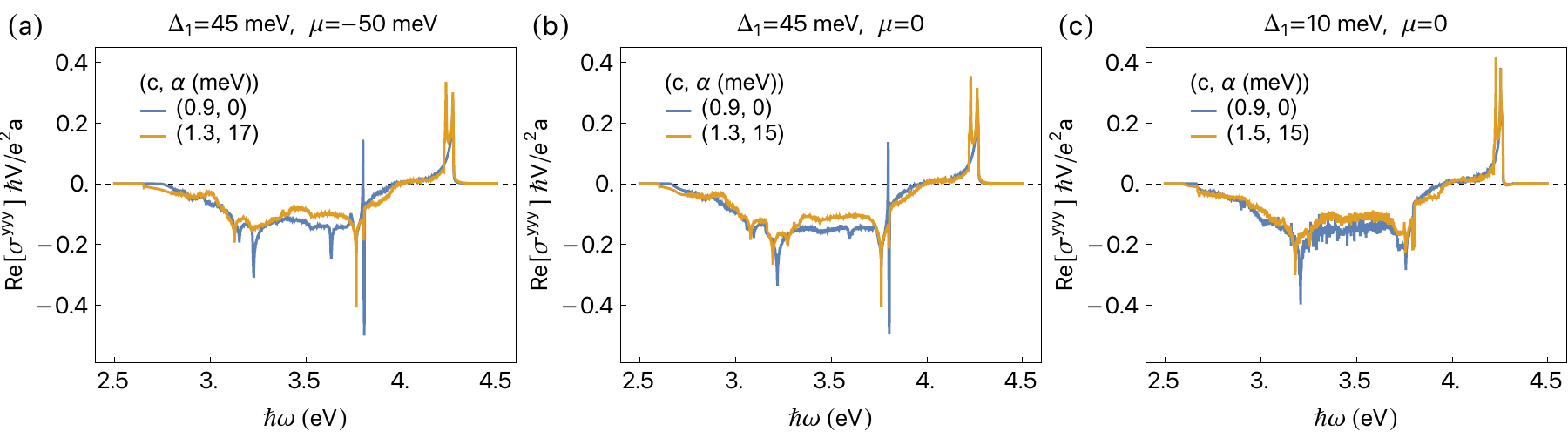}
    \caption{$\rm{Re}[\s^{yyy}(\w)]$ at higher frequencies in the trivial (blue) and TRITOPS (orange) phases for different values of $\D_1$ and $\mu$. There is no significant change with changing either $\mu$ (a-b) or $\D_1$ (b-c).}
    \label{fig:sigmayyy3}
\end{figure*}

We study the second order, Eq.~\eqref{eq:sigma2}, and the first order, Eq.~\eqref{eq:sigma1}, optical responses in two different frequency ranges--one on the order of the superconducting gap, and the other much above it. The focus on the low energy window is motivated by the size of the superconducting terms in Eq.~\eqref{eq:HBdG} and hence captures the role of transitions between particle and hole copies of bands near the Fermi energy. These transitions rely on non-zero superconducting pairing terms and their parity. On the other hand, for the higher energy window, the optical response arises mainly from transitions between different single-particle bands. The band structures for the normal and the superconducting Hamiltonians are shown in Fig.~\ref{fig:bands}.

The real and imaginary parts of the first-order conductivity, Eq.~\eqref{eq:sigma1}, are shown in Fig.~\ref{fig:sigmayy} for $\D_1=45$ meV and chemical potential, $\mu=-50$ meV in the two frequency windows. The electric field of the linearly polarized light is taken to be in the $y$-direction. We see no features indicating a phase transition in the linear conductivity. For our system, $\s^{xx}=\s^{yy}$ whereas $\s^{xy}=0$ due to the presence of time-reversal symmetry.

The second-order conductivity results obtained from Eq.~\eqref{eq:sigma2} are shown in Figs.~\ref{fig:sigmayyy1}, \ref{fig:sigmayyy2}, and \ref{fig:sigmayyy3}. Unlike the linear conductivity, the $\s^{yyy}$ (again, the electric field is taken to be in the $y$-direction for linearly polarized light and one has $\s^{yxx}=-\s^{yyy}$ due to three-fold rotation symmetry) component of the second-order conductivity shows distinct signatures when approaching the nodal phase from the trivial and the TRITOPS phases in the low-frequency regime ($\w \lesssim \D_1$). In particular, Re$[\s^{yyy}]$ shows a zero-frequency divergence at the trivial to nodal phase transition, as shown in Figs.~\ref{fig:sigmayyy1} (a-c) and \ref{fig:sigmayyy2} (a) for different values of $\D_1$ and chemical potential. However, at the TRITOPS to nodal phase transition, there is an opposite sign zero-frequency divergence as shown in Figs.~\ref{fig:sigmayyy1} (d-f) and \ref{fig:sigmayyy2} (b,c). We also find that close to the two phase boundaries, the sign of the low-frequency photogalvanic response in the trivial and TRITOPS phases matches with that of the divergence. For the trivial phase this behavior is captured in Figs.~\ref{fig:sigmayyy1} (a-c) whereas for the TRITOPS phase it is shown in Figs.~\ref{fig:sigmayyy1} (d-f). This behavior remains qualitatively unchanged when we decrease $\D_1$ from 100 meV to 45 meV, and finally to 10 meV. We note that the behavior of $\s^{yyy}$ across the nodal-TRITOPS phase boundary does not depend on whether the transition was driven by $c$ or $\A$, making it a useful signature of the phase transition itself. 

We also note that $\s^{yyy}$'s approach to the divergences at the nodal phase boundaries is different from within the nodal phase and from the outside. While the trivial and TRITOPS phase conductivities' approach to their respective divergences is gradual, the nodal phase conductivity shows rapid sign change close to the phase boundaries. For the nodal-TRITOPS phase boundary, this can be seen in Figs.~\ref{fig:sigmayyy2} (b) and (c). The nodal phase conductivity develops two peaks [marked as A and B in Fig.~\ref{fig:sigmayyy2} (c)]. Peaks A and B get closer to $\w=0$ line while becoming increasingly negative and positive, respectively, as the system approaches the transition point. As the system moves closer to the phase boundary, Peak A (now indistinguishable from Re[$\s^{yyy}(0;0,0)$]) rapidly moves up and merges with B to give the divergence at the nodal-TRITOPS phase boundary. More details about the behavior of Re[$\s^{yyy}(0;0,0)$] is given in Appendix~\ref{appendix:0divergence}. We emphasize that just like the TRITOPS phase conductivity peak C, the approach of peak B to its divergence is much more gradual compared to A.

We also considered the effect of doping on the second-order conductivity for the $\D_1=45$ meV case by taking $\mu=-50$ meV. Our results are shown in Fig.~\ref{fig:sigmayyy2}. A finite value of $\mu$ increases the critical values of $c,\,\A$ required for either phase transition, as shown in Table~\ref{tab:finitmu}. However, the behavior of $\s^{yyy}$ around the shifted phase transition points remains unchanged from the $\mu=0$ case seen in Fig.~\ref{fig:sigmayyy1}.

Finally, Fig.~\ref{fig:sigmayyy3} shows that the high frequency photogalvanic response remains unaffected across the phase transitions for different values of $\D_1$ and $\mu$ which makes it useful as a reference point for analyzing the relative sign of the low-frequency divergences and responses around them. We find that the high frequency response is non-zero for $2.6\text{ eV}\lesssim\w\lesssim4.3\text{ eV}$. For most of this window, the response is of the same sign as the divergence at the trivial-nodal phase transition point. Since the divergence at the nodal-TRITOPS phase transition is of the opposite sign, this observation can be used to distinguish the trivial and TRITOPS phases in experiments.  Whether this is a specific feature of the Hamiltonian we study or is true more generally would require different Hamiltonians with the same phase diagram to be studied.  Our main purpose here is to show that for 1H-TaS$_2$ and 4Hb-TaS$_2$, for which our model is relevant~\cite{margalit2021}, these features can be used to identify the phase and provide complimentary information to other experimental studies~\cite{silber2022chiral, nayak2021evidence, ribak2020chiral}.  We hope our work will help inspire experimental groups to undertake this challenge.

% ----------------------------------------------------------------------------------------
\section{Conclusions}
\label{sec:Conclusions}
We have presented a thorough--both low frequency and high frequency regimes--study of the second-order DC response in 1H-Ta$\rm{S}_2$, of which the candidate topological superconducting 4Hb-Ta$\rm{S}_2$ compound is partially built \cite{nayak2021evidence,ribak2020chiral}.  (The 4Hb-Ta$\rm{S}_2$ compound is composed of alternating layers of 1H-Ta$\rm{S}_2$ and 1T-Ta$\rm{S}_2$.) Based on the ratio of inter- and intra-orbital pairing amplitudes and the presence of Rashba spin-orbit coupling permitted by inversion symmetry breaking, the system is known to exist in one of three phases: trivial, nodal, and TRITOPS. We have mapped out the phase boundaries by analyzing the gap closing and reopening at the Fermi level. With the phase diagram in hand, we have numerically calculated the first and second-order conductivities around these transition points. Our results indicate that the transitions from trivial to nodal phases and from nodal to TRITOPS phases are each characterized by a zero-frequency divergence in the photogalvanic response but with opposite signs. No signature of the superconducting phase of the system is observed in the linear response. This makes the photogalvanic response an effective probe to distinguish the superconducting phases of 1H-Ta$\rm{S}_2$ and potentially the closely related 4Hb-Ta$\rm{S}_2$ compound. 

The topological phase transition in 1H-Ta$\rm{S}_2$ depends on the extent of parity mixing in the superconducting pairing and the strength of the Rashba spin-orbit coupling. The parity mixing relies on the symmetry aspects of the substrate (e.g., broken inversion symmetry), and hence the proximity/coupling strength to a substrate can be used a knob to control the ratio of the two opposite parity components of the superconducting order parameter. The Rashba spin-orbit coupling also arises from broken inversion symmetry, and thus can be possibly varied either by substrate engineering or by applying an out-of-plane electric field, allowing independent control of the parity mixing of the superconducting order parameter and the Rashba spin-orbit coupling. 

The superconductivity in 1H-Ta$\rm{S}_2$ is very robust to an in-plane magnetic field which may serve as another knob to modify topological properties~\cite{seshadri2022josephson}. In future studies it would be interesting to determine if the quantum phase transitions in the presence of an in-plane magnetic field would also lead to some unique signatures in the second-order DC response. 

Further theoretical studies of distinct experimental signatures of topological superconductivity in different measurements can be used to more clearly identify whether a given material indeed supports topological superconductivity, rather than relying on one class of measurements alone. Given the controversy around purported topological superconductors, multiple measurement signatures of topology in superconductors is highly desirable.

% ----------------------------------------------------------------------------------------
\section{Acknowledgements}
% ----------------------------------------------------------------------------------------
We acknowledge funding from the National Science Foundation through the
Center for Dynamics and Control of Materials: an
NSF MRSEC under Cooperative Agreement No. DMR-
1720595 and DMR-2114825. G.A.F. acknowledges additional support the Alexander von Humboldt Foundation.

% ----------------------------------------------------------------------------------------
\appendix
% ----------------------------------------------------------------------------------------
\section{Details of Model}\label{appendix:model}
\subsection{TMD monolayer Hamiltonian}

The Hamiltonian for the transition metal dichalcogenide monolayer without superconductivity is given by
\begin{equation}
H_0(\mathbf{k}) = E + \sum_{j=1}^6 R_j e^{i\mathbf{R}_j\cdot\mathbf{k}} + \sum_{j=1}^6 S_j e^{i\mathbf{S}_j\cdot\mathbf{k}} + \sum_{j=1}^6 T_j e^{i\mathbf{T}_j\cdot\mathbf{k}},
\end{equation}
where $\mathbf{R}_j$, $\mathbf{S}_j$ and $\mathbf{T}_j$ are the first, second and third-nearest neighbor lattice vectors, respectively, and $R_j$, $S_j$, and $T_j$ are the corresponding hopping matrices. $E$, $R_1$, $S_1$, $T_1$ are defined as,
\begin{equation}
E = \s_0 \otimes
    \begin{bmatrix}
        \epsilon_0 - \mu & 0 & 0\\
        0 & \epsilon_1-\mu & 0\\
        0 & 0 & \epsilon_2-\mu
    \end{bmatrix}
    + \s_z \otimes 
    \begin{bmatrix}
        0 & 0 & 0\\
        0 & 0 & i\lambda_{SO}\\
        0 & -i\lambda_{SO} & 0\\
    \end{bmatrix},
\end{equation}
\begin{equation}
    R_1 = \s_0 \otimes 
    \begin{bmatrix}
        t_0 & -t_1 & t_2\\
        t_1 & t_{11} & -t_{12}\\
        t_2 & t_{12} & t_{22}
    \end{bmatrix},
\end{equation}
\begin{equation}
    S_1 = \s_0 \otimes 
    \begin{bmatrix}
        r_0 & r_2 & -\frac{1}{\sqrt{3}}r_2\\
        r_1 & r_{11} & r_{12}\\
        -\frac{1}{\sqrt{3}}r_1 & r_{12} & (r_{11} + \frac{2}{\sqrt{3}}r_{12})
    \end{bmatrix},
\end{equation}
\begin{equation}
    T_1 = \s_0 \otimes 
    \begin{bmatrix}
        u_0 & -u_1 & u_2\\
        u_1 & u_{11} & -u_{12}\\
        u_2 & u_{12} & u_{22}
    \end{bmatrix},
\end{equation}
whereas the remaining hopping matrices can be generated via the following:
\begin{equation}
    C_3 = 
    \begin{bmatrix}
        e^{-i\frac{\pi}{3}} & 0\\
        0 & e^{i\frac{\pi}{3}}
    \end{bmatrix}
    \otimes 
    \begin{bmatrix}
        1 & 0 & 0\\
        0 & -\frac{1}{2} & \frac{\sqrt{3}}{2}\\
        0 & -\frac{\sqrt{3}}{2} & -\frac{1}{2}
    \end{bmatrix}
\end{equation}
\begin{equation}
  \begin{array}{ >{\arraybackslash$} p{2.4cm} <{$} >{\arraybackslash$} p{2.4cm} <{$} >{\arraybackslash$} p{2.4cm} <{$}}
    R_2 = C_3^\dagger R_1^\dagger C_3 & S_2 = C_3^\dagger S_1^\dagger C_3 & T_2 = C_3^\dagger T_1^\dagger C_3 \\
    R_3 = C_3 R_1 C_3^\dagger & S_3 = C_3 S_1 C_3^\dagger & T_3 = C_3 T_1 C_3^\dagger\\
    R_4 = R_1^\dagger & S_4 = S_1^\dagger & T_4 = T_1^\dagger\\
    R_5 = C_3^\dagger R_1 C_3 & S_5 = C_3^\dagger S_1 C_3 & T_5 = C_3^\dagger T_1 C_3\\
    R_6 = C_3 R_1^\dagger C_3^\dagger & S_6 = C_3 S_1^\dagger C_3^\dagger & T_6 = C_3 T_1^\dagger C_3^\dagger
\end{array}  
\end{equation}

\begin{table}[ht!]
\caption{Values of the hopping parameters taken from~\cite{margalit2021}.}
\begin{ruledtabular}
\begin{tabular}{cccccc}
 $t_0$ & $t_1$ & $t_2$ & $t_{11}$ & $t_{12}$ & $t_{22}$\\
 -0.1917 & 0.4057 & 0.4367 & 0.2739 & 0.3608 & -0.1845 \\
 $r_0$ & $r_1$ & $r_2$ & $r_{11}$ & $r_{12}$ & $r_{22}$\\
 0.0409 & -0.069 & 0.0928 & -0.0066 & 0.1116 & 0. \\
 $u_0$ & $u_1$ & $u_2$ & $u_{11}$ & $u_{12}$ & $u_{22}$\\
 0.0405 & -0.0324 & -0.0141 & 0.1205 & -0.0316 & -0.0778 \\
 $\epsilon_0$ & $\epsilon_1$ & $\epsilon_2$ & $\lambda_{SO}$ & &\\
 1.6507 & 2.5703 & 2.5703 & 0.1713 &  &  
\end{tabular}
\end{ruledtabular}
\label{tab:parameters}
\end{table}

% ----------------------------------------------------------------------------------------
\section{Zero frequency divergence in Re[$\s^{yyy}$]} \label{appendix:0divergence}

We set $\w_1=-\w_2=0$ and look at the integrand of $\s^{yyy}(0;0,0)$ from Eq.~\eqref{eq:sigma2}. After some simplification we obtain,
\begin{align}
\begin{split}
    & \frac{1}{2\eta^2} \Bigg[\sum_a \frac{1}{2}J_{aa}^{yyy}f_a \\
    &\quad + \sum_{a,b} \frac{1}{2}J_{ab}^{yy}J_{ba}^y f_{ab} \Bigg( \frac{2}{i\eta -E_{ba}} - \frac{1}{2i\eta +E_{ba}} \Bigg) \\
    &\quad + \sum_{a,b,c} \frac{J_{ab}^y J_{bc}^y J_{ca}^y}{2i\eta-E_{ba}} \Bigg(\frac{ f_{ac}}{i\eta-E_{ca}}-\frac{f_{cb}}{i\eta -E_{bc}}\Bigg)\Bigg].
\end{split}
\label{eq:0div}
\end{align}
It is easily seen that the integrand is real.
To simplify it further, we set $T=0$ K, so $f_a=1-\theta(E_a)$ and $f_{ab}=\theta(E_b)-\theta(E_a)$, where $\theta(x)$ is the Heaviside step function. The structure of the Bogoliubov-de Gennes Hamiltonian is such that it has symmetric eigenvalues with respect to zero energy. If the eigenvalues are sorted in ascending order at each k-point (bands labeled 0-11), then only bands 5 and 6 are important for capturing the divergence since they are the bands closest to zero energy. Picking out terms involving these bands from Eq.~\eqref{eq:0div}, we get 
\begin{figure}[ht!]
    \centering
    \includegraphics[scale=0.55]{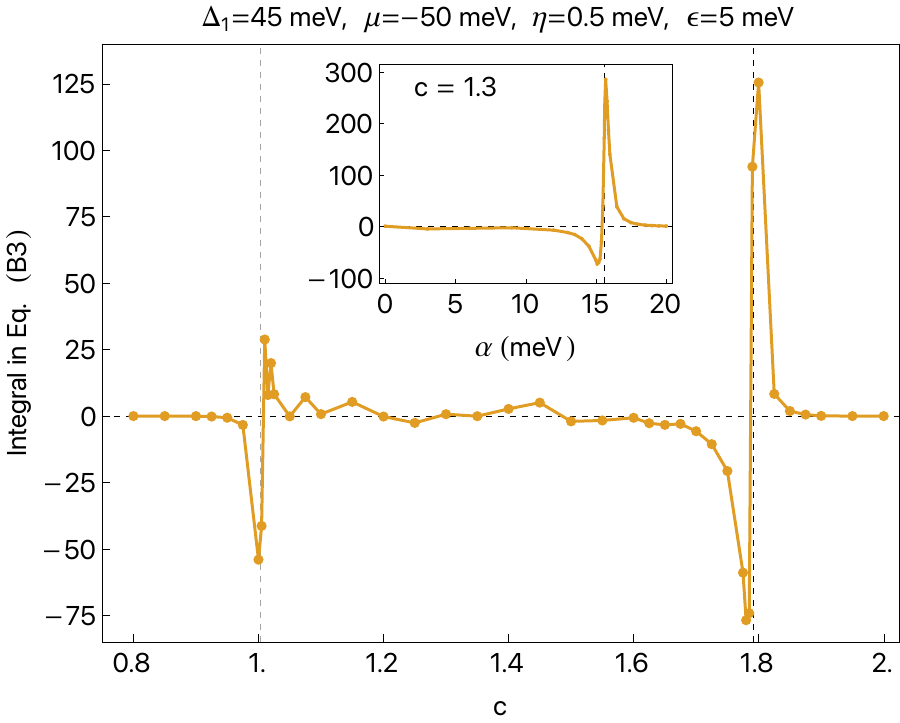}
    \caption{Behavior of the integral in Eq.~\eqref{eq:0integral} across the trivial-nodal ($c=1.0019486$) and nodal-TRITOPS ($c=1.7917638$) phase transition points. Inset: Behavior of the same integral across the nodal-TRITOPS phase boundary when the transition is driven by $\A$. Note that for $c=1.3$, $\A_c=15.6259$ meV.}
    \label{fig:divergence}
\end{figure}
\begin{align}
\begin{split}
    & \frac{1}{2\eta^2} \Bigg[ \frac{1}{2}J_{55}^{yyy} + \text{Re}\Bigg[J_{56}^{yy}J_{65}^y \Bigg( \frac{2}{i\eta -2E_6} - \frac{1}{2i\eta +2E_6} \Bigg)\Bigg] \\
    &\quad + \text{Re}\Bigg[\frac{J_{56}^y J_{65}^y}{i\eta+E_{6}}\,\frac{J_{66}^y-J_{55}^y}{i\eta+2E_{6}}\Bigg] + J_{56}^y J_{65}^y \Bigg(\frac{J_{66}^y-J_{55}^y}{\eta^2+4E_{6}^2}\Bigg) \\
    &\quad + \frac{J_{56}^y J_{65}^y}{2} \Bigg(\frac{J_{66}^y-J_{55}^y}{(i\eta+2E_{6})^2}+\frac{J_{66}^y-J_{55}^y}{(i\eta-2E_{6})^2}\Bigg) \Bigg].
\end{split}
\end{align}
One has $J_{55}^{yyy}(\kv)=-J_{55}^{yyy}(-\kv)$ and $J^{y}_{55}(\kv)=J^{y}_{66}(\kv)$. The first term contributes nothing when integrated (note $E_a(\kv)=E_a(-\kv)$), whereas the third, fourth, and fifth terms are zero. Thus, only the second term remains
\begin{align}
    & \int_{E_6(\kv)<\epsilon} \frac{\dd[2]{k}}{8\pi^2\eta^2} \text{Re}\Bigg[J_{56}^{yy}J_{65}^y \Bigg( \frac{2}{i\eta -2E_6} - \frac{1}{2i\eta +2E_6} \Bigg)\Bigg],
\label{eq:0integral}
\end{align}
where $\epsilon$ is a small cutoff (staying close to the Fermi level where the low-energy approximation is reliable). Near the node there is a Dirac-like dispersion, and small $\epsilon$ keeps one within the linear regime. The divergence of the integrand is numerically shown in Fig.~\ref{fig:divergence} as a function of the parameter $c$ and $\A$ which control the phase of the superconductor. The structure of the divergences is consistent with the conductivity plots in Fig.~\ref{fig:sigmayyy2}, providing a clearer picture of its origin.

\bibliography{ref.bib}
\end{document}